\documentclass[conference]{IEEEtran}
\IEEEoverridecommandlockouts
\usepackage{cite}
\usepackage{amsmath,amssymb,amsfonts}
\usepackage{algorithmic}
\usepackage{graphicx}
\usepackage{textcomp}
\def\BibTeX{{\rm B\kern-.05em{\sc i\kern-.025em b}\kern-.08em
    T\kern-.1667em\lower.7ex\hbox{E}\kern-.125emX}}
    
\usepackage{balance}
\usepackage[linesnumbered,ruled,vlined]{algorithm2e}

\SetCommentSty{mycommfont}
\usepackage[dvipsnames]{xcolor}
\usepackage[caption=false,font=footnotesize,labelfont=sf,textfont=sf]{subfig}
\usepackage{textcomp}
\usepackage{stfloats}
\usepackage{multirow}
\usepackage{etoolbox}
\usepackage{pgfplots, pgfplotstable}
\usepackage[font=small,labelfont=bf]{caption}
\usepgfplotslibrary{groupplots}
\usepackage{tikz}
\usetikzlibrary{matrix, positioning, patterns, shapes, arrows}
\usepackage{tikzscale}
\usepgfplotslibrary{statistics}
\pgfplotsset{compat=newest}% <- added, current version is 1.13
\usepgfplotslibrary{colorbrewer}

\usepackage{lipsum}

\usepackage{microtype}

\makeatletter
\def\ps@IEEEtitlepagestyle{
\def\@oddfoot{\mycopyrightnotice}
\def\@evenfoot{}
}
\def\mycopyrightnotice{
{\footnotesize 978-1-6654-8524-1/22/\$31.00~\copyright2022 IEEE\hfill} % <--- Change here
\gdef\mycopyrightnotice{}
}

\graphicspath{{figs/}}
\newcommand{\fixme}[2]{\ifx&#2&{\color{red}#1}\else{\color{red}FIXME\{}#1{\color{red}\}}\footnote{{\color{red}#2}}\PackageWarning{Fixme}{#1: #2}\fi}

\begin{document}

\title{Multi-Factor Pruning for Recursive Projection-Aggregation Decoding of RM Codes
%{\footnotesize \textsuperscript{*} }
%\thanks{This paper has been accepted for SiPS 2022.}
}

\author{
\IEEEauthorblockN{Marzieh Hashemipour-Nazari and Kees Goossens and Alexios Balatsoukas-Stimming}
\IEEEauthorblockA{\textit{Department of Electrical Engineering} \\
\textit{Eindhoven University of Technology}\\
Eindhoven, The Netherlands \\
\{s.m.hashemipour.nazari,k.g.w.goossens,a.k.balatsoukas.stimming\}@tue.nl}}
%\and
%\IEEEauthorblockN{Kees Goossens}
%\textit{ Eindhoven University of Technology}\\
%Eindhoven, The Netherlands\\
%email address or ORCID}
%\and
%\IEEEauthorblockN{Alexios Balatsoukas-Stimming}
%\IEEEauthorblockA{\textit{Department of Electrical Engineering} \\
%\textit{ Eindhoven University of Technology}\\
%Eindhoven, The Netherlands\\
%email address or ORCID}
%}

\maketitle

\begin{abstract}
The recently introduced recursive projection aggregation (RPA) decoding method for Reed-Muller (RM) codes can achieve near-maximum likelihood (ML) decoding performance. However, its high computational complexity makes its implementation challenging for time- and resource-critical applications. In this work, we present a complexity reduction technique called multi-factor pruning that reduces the computational complexity of RPA significantly. Our simulation results show that the proposed pruning approach with appropriately selected factors can reduce the complexity of RPA by up to $92\%$ for $\text{RM}(8,3)$ while keeping a comparable error-correcting performance.
\end{abstract}


\section{Introduction}
\label{section:intro}

Reed-Muller (RM) codes are linear block codes with a recursive structure introduced in 1954 by Muller~\cite{Muller1954}. Immediately after, Reed\cite{Reed1954} introduced the first decoder for RM codes based on majority voting, which could correct bit errors in a codeword up to half of the minimum distance of the corresponding code. Since then, several efficient decoding methods for RM codes have been published~\cite{Betextquotesingleery1986,Ashikhmin1996,green1966serial,Sidel1992,Dumer2004,Sakkour2005,Dumer2006a,Dumer2006}. Recently, there has been growing interest in RM codes as they have been shown to be capacity-achieving under the maximum-likelihood (ML) decoding for general symmetric channels~\cite{Arikan2009, Hussami2009, Kudekar2017, Abbe2015}. However, since the complexity of ML decoding scales exponentially with the blocklength, ML decoding is not a practically feasible decoding method. Therefore, considerable effort has been put into proposing efficient near-ML decoding algorithms for RM codes with tractable complexity, which is especially attractive for short blocklength applications~\cite{Tonnellier2021}.

A wide variety of such proposed algorithms are taking advantage of the recursive structure of the RM codes to achieve near-ML decoding. For example, the recursive list decoder proposed in \cite{Dumer2006} provides efficient near-ML decoding performance for RM codes with reasonable list size. Nevertheless, similarly to all list-based decoders, it is challenging to achieve low-latency decoding even with a relatively small list size due to the required path selection operation that cannot be readily parallelized. Moreover, due to the similarity of RM codes to polar codes, the successive-cancellation (SC) decoding~\cite{Arikan2009} and SC list (SCL)~\cite{Tal2011} decoding algorithms for polar codes are able to decode RM codes recursively with reasonable complexity.

Recently, a recursive algorithm called \emph{recursive projection-aggregation} (RPA)~\cite{Ye2020} decoding was proposed for decoding RM codes. RM codes decoded with the RPA algorithm achieve near-ML decoding performance, outperforming polar codes with the same blocklength and rate under SCL decoding even with large list sizes. On a high level, the RPA decoder generates many smaller codewords by combining elements of the received vector and then decodes the generated codewords recursively to aggregate them into a reliable decision for the received vector.
The authors of~\cite{Ye2020} mention that the computational complexity of RPA decoding scales like $O(n^r \log n)$ where $n$ is the blocklength and $r$ is the order of the RM code, making RPA decoding impractical for RM codes with large order $r$. 

Several algorithms have been published to reduce the complexity of the RPA algorithm. For example, simplified RPA~\cite{Ye2020} and collapsed projection-aggregation (CPA)~\cite{Lian2020} reduce the number of recursive levels and project the received vectors directly onto smaller codes, thus reducing the number of projections. However, both simplified RPA and CPA lead to high-complexity projection and aggregation steps. The authors of  
\cite{Fathollahi2021} introduced a variant of the RPA decoder called sparse RPA that uses a small random subset of the projections, thus reducing the complexity of the RPA algorithm. However, the error-correcting performance of SRPA degrades significantly as the sparsity increases. To address this, the authors also proposed \mbox{k-SRPA} which employs $k$ sparse decoders to improve the error-correcting performance of single SRPA. A cyclic redundancy check (CRC) and Reed's algorithm are used to select the most reliable candidate among all candidates decoded from the $k$ individual SRPA decoders. Unfortunately, the \mbox{k-SRPA} algorithm causes a reduction in the effective code rate due to the added CRC as well as hardware overhead due to the required randomization of projections, multiple control units, memories, and data paths for multiple decoders. The work of \cite{JiaJie2021} proposed syndrome-based early stopping techniques along with a scheduling scheme to reduce the complexity of the RPA algorithm called reduced complexity RPA. Reduced complexity RPA is still challenging due to the computational overhead of syndrome checks and the variable decoding latency. 

%Therefore, the RPA decoding method and its simplifications are still challenging to implement on hardware as they need large computational resources, resulting in large circuits. 
In this work, we further optimize RPA decoding with the ultimate goal of making it feasible for hardware implementation. More specifically, we propose a multi-factor pruning method that is inspired by observations we made about the relative importance of projections at various recursion levels based on simulations.
Our method prunes the RPA decoder with a varying level of aggressiveness for different iterations and recursion levels.
% This method uses different \fixme{pruning parameters}{You have not explained what ``pruning parameters'' are.} regarding the \fixme{running iteration}{What is the ``running iteration''?} and recursion level in the RPA algorithm. Consequently, \fixme{as the proposed pruning factors vary in recursive layers,}{This feels repeated, say what you want to say more directly.} 
Our results show a computational complexity reduction of $92\%$ compared to the baseline RPA algorithm~\cite{Ye2020} and $38$\% to $77$\% compared to the works of \cite{JiaJie2021} and \cite{Fathollahi2021} with no degradation in the error-correcting performance.
Moreover, in contrast to the works of \cite{Fathollahi2021} and \cite{JiaJie2021}, our proposed pruning method can translate to a simpler hardware implementation.

The remainder of this paper is organized as follows. In Section \ref{sec:background}, we give the preliminaries and background, including a description of RM codes as well as the RPA algorithm and we explain the issues with previously proposed pruning methods in more detail. In Section~\ref{sec:delta_RPA}, we present our proposed multi-factor pruning method. In Section \ref{sec:results}, we present error-correcting performance simulation results and computational complexity comparisons with existing pruning methods. Finally, we conclude the paper in Section~\ref{sec:conclusion}.

\section{Preliminaries and Background}
\label{sec:background}
\subsection{Reed-Muller codes}
Reed-Muller codes are linear block codes denoted by $\text{RM}(m,r)$, where $n = 2^m$ is the blocklength, and $r$ is the order of the code.
Similar to the other linear block codes, RM codes are defined by a $k \times n$ generator matrix $\mathbf{G}_{(m,r)}$, where $k=\sum _{i=0}^r \binom{n}{i}$ and the rate of the code is $R= \frac{k}{n}$. A binary vector $\mathbf{u}$ with $k$ information bits is encoded into a binary vector $\mathbf{c}$ belonging to the $\text{RM}(m,r)$ code as follows:
\begin{equation} \label{eq:code}
\mathbf{c}= \mathbf{u}\mathbf{G}_{(m,r)}.
\end{equation}
The generator matrix $\mathbf{G}_{(m,r)}$ of $\text{RM}(m,r)$ code is obtained from the $n^\text{th}$ Kronecker power of $\mathbf{F}=\begin{bmatrix}
  1 & 1\\ 
  0 & 1
\end{bmatrix}$ by selecting the rows with a Hamming weight of at least $2^{m-r}$. The resulting generator matrix $\mathbf{G}_{(m,r)}$ has a universal and recursive structure defined as: 
\begin{equation}\label{eq:GMtx}
\mathbf{G}_{(m,r)}= \begin{bmatrix}
  \mathbf{G}_{(m-1,r)} & \mathbf{G}_{(m-1,r)}\\ 
  \mathbf{0} & \mathbf{G}_{(m-1, r-1)}
\end{bmatrix},
\quad
\mathbf{G}_{(1, 1)}= \mathbf{F}.
\end{equation}
In addition, a codeword $\mathbf{c}\in \text{RM}(m,r) $ can be be mapped to an $m$-variate polynomial with degree less than or equal to $r$ in a vector space $\mathbb{E}:= \mathbb{F}^{m}_{2}$.
Consequently, any  coordinate of the codeword $\mathbf{c}$ are indexed by an $m$-bit binary vector $z \in \mathbb{E}$,i.e., $\mathbf{c}=\left(c(z),z\in \mathbb{E} \right)$~\cite{Abbe2020}.

\subsection{Coset-based projections}
\label{sec:coset}
Let $\mathbb{B}$ be an $s$-dimensional subspace of $\mathbb{E}$. There are $2^{m-s}$ different cosets $T$ for $\mathbb{B}$ making the quotient space $\mathbb{E}/ \mathbb{B}$:
\begin{equation}\label{eq:t}
\mathbb{E}/ \mathbb{B}  = \lbrace T:= z+\mathbb{B}, z\in \mathbb{E} \rbrace.
\end{equation}
Moreover, the projection of $\mathbf{c}=\left(c(z),z\in \mathbb{E} \right)$ on the cosets of an $s$-dimensional subspace $\mathbb{B}$ is defined as:
\begin{equation}
\mathbf{c}_{/ \mathbb{B}}=\operatorname{Proj}(\mathbf{c}, \mathbb{B}):=\left(c_{/ \mathbb{B}}(T), T  \in \mathbb{E} / \mathbb{B} \right),
\end{equation} 
where $\mathbf{c}_{/ \mathbb{B}} = \left(c_{/ \mathbb{B}}(T):= {\oplus}_{z\in T}c(z)\right)$ is a binary vector generated by summing up all coordinates of $\mathbf{c}$ indexed by the elements of each coset $T\in \mathbb{E}/ \mathbb{B}$. 
The work of~\cite{Ye2020} proved that $c_{/ \mathbb{B}}$ is a codeword of $\text{RM}(m-s,r-s)$ if $\mathbf{c}$ is a codeword of $\text{RM}(m,r)$. The RPA decoding algorithm, which is the focus of this paper, exploits this property of RM codes using \mbox{one-dimensional} subspaces, i.e., $s=1$.

%
%There are $n-1$ one-dimensional subspaces for binary vector space $\mathbb{E}$, providing $n-1$ projections for vector $\mathbf{y}$ based on their corresponding quotient space such that:
%
%\begin{equation}\label{eq:proj}
%y_{/ \mathbb{B}_i}=\operatorname{Proj}(y, \mathbb{B}_i):=\left(y_{/ \mathbb{B}_i}(T_t), T_t \in \mathbb{E} / \mathbb{B}_i\right).
%\end{equation}  
%$y_{/ \mathbb{B}_i}$ is a binary vector obtained by taking the sum over every two coordinates of $\mathbf{y}$ indexed by $z \in T_{t}$ of the $2^{m\text{-}1}$ cosets of $\mathbb{B}_i$.
%It's proved that if $c\in RM(m,r)$ results in $c_{/\mathbb{B}}\in RM(m\text{-}1,r\text{-}1)$ \cite{Ye2020}.

\subsection{Recursive projection aggregation (RPA) Decoding}
\label{sec:RPA}
%------RPA algorithm ----------------------
\begin{figure}[t]
  \centering
  \centerline{\includegraphics[width=0.525\textwidth]{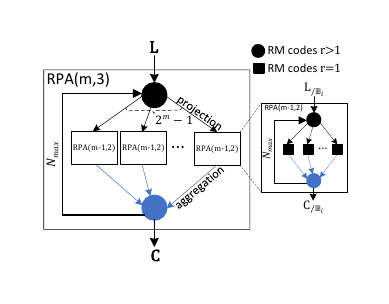}}
  \vspace{-1cm}
  \caption{\small RPA decoding for $r=3$ RM codes.}
  \label{fig:RPA}
  \vspace*{-1em}
\end{figure} 
%---------------------------------------------

Let $W:\{0,1\} \rightarrow \mathcal{W}$ denote a binary-input memoryless channel, where $\mathcal{W}$ is the output alphabet.
The log-likelihood ratio (LLR) of a channel output $y\in \mathcal{W} $, denoted by $L(y)$, is defined as:
\begin{equation}\label{eq:LLR}
{ L(y)}:=\ln \left(\frac{W(y \mid 0)}{W(y \mid 1)}\right).
\end{equation}
%However, hard-decision RPA takes the binary value $\mathbf{x}$ as the input, where 
%\begin{equation}\label{eq:hard-decision}
%x=\frac{1-\text{sign}(\boldsymbol{L})}{2}.
%\end{equation}
Soft-decision RPA decoding uses the LLR values of the channel output vector $\mathbf{y}$, denoted by the vector $\mathbf{L}$, as its input. In particular, the RPA decoding algorithm decodes $\mathbf{L}$ in three steps: projection, recursive decoding, and aggregation.

As Fig.~\ref{fig:RPA} shows, in the \textit{projection} step the RPA algorithm generates $n-1$ vectors $\mathbf{L}_{/\mathbb{B}_i}$ with length of $2^{m-1}$   by projecting $\mathbf{L}$ onto $n-1$ one-dimensional subspaces of the vector space $\mathbb{E}$. 
Let $\mathbb{B}_i=[0,i]$ be the $i$-th one-dimensional subspace of $\mathbb{E}$. There are $2^{m-1}$ different cosets $T$ for $\mathbb{B}_i$. Therefore, the quotient space $\mathbb{E}/ \mathbb{B}_i$ is built based on~\eqref{eq:t}.
%\begin{equation}\label{eq:t}
%\mathbb{E}/ \mathbb{B}_i  = \lbrace T_{t}:= [t,i \oplus t], t\in \mathbb{E} \rbrace.
%\end{equation}
There are $n-1$ one-dimensional subspaces for the binary vector space $\mathbb{E}$ that provide $n-1$ projections for channel output vector $\mathbf{L}$ given their corresponding quotient space such that:

\begin{equation}
\mathbf{L}_{/ \mathbb{B}_i}=\operatorname{Proj}(\mathbf{L}, \mathbb{B}_i):=\left({L}_{/ \mathbb{B}_i}(T), T \in \mathbb{E} / \mathbb{B}_i\right),
\end{equation}  
where
\begin{equation}
{L}_{/ \mathbb{B}_i}(T)=2\tanh^{-1}\left(\prod_{z\in T} \tanh\left(\frac{L(z)}{2}\right) \right).
\end{equation}  

In the \textit{recursive decoding} step, each projected vector $\mathbf{L_{/\mathbb{B}_i}}$ is recursively decoded by RPA for $\text{RM}(m{-}1,r{-}1)$ until first-order RM codes are reached. At this point, the RPA algorithm uses an efficient decoder based on the fast Hadamard transform (FHT) to decode the first-order RM codes~\cite{Betextquotesingleery1986}.

In the final \textit{aggregation} step a per-coordinate average LLR is calculated from all the decoded codewords obtained from the previous step as follows:
\begin{equation}\label{eq:agg}
{\hat{L}}(z)=\frac{1}{n-1}\sum_{i=1}^{n-1}\left(1-2\hat{c}_{/ \mathbb{B}_i}(z+\mathbb{B}_i)\right)L(z+z_i),
\end{equation}
where $\hat{c}_{/ \mathbb{B}_i}$ is the decoded binary codeword from recursive decoding, and $\hat{L}(z)$ is the LLR value of $z^{\text{th}}$ index of the estimated vector  $\hat{\mathbf{c}}$ for the transmitted codeword $\mathbf{c}$.

The RPA decoding algorithm repeats the above steps up to $N_{\max}$ times or until the output converges to the input as:  
\begin{equation}\label{eq:earlystop}
|\hat{L}(z)-{L}(z)|<\theta|L(z)|,\forall z\in\mathbb{E},
\end{equation}
where $\theta$ is a small constant called the exiting threshold. %\cite{Ye2020} set $\theta=0.05$. 
%The complexity of FHT-based first-order decoder (FOD) used in the most internal recursion level of the RPA decoder scales like $O(n\log n)$, where $n$ is the length of the FHT input. 
The total number of FHT-based first-order decodings (FODs), which we use as a proxy for the computational complexity,\footnote{We ignore the complexity of the projection and aggregation steps for simplicity because the overall complexity of RPA decoding is dominated by the FOD operations.} is:
\begin{align}\label{eq:rpa_comp}
\lambda_{\text{RPA}}=N_{\max}^{r-1}\prod_{i=1}^{r-1}\left(2^{m-i-1}-1\right), 
\end{align}
as mentioned in \cite{JiaJie2021}. 

%Considering $\lambda_{\text{RPA}}$ as computational complexity indicator for RPA decoder and setting $N_{\max}=\lceil{\frac{m}{2}}\rceil$ as in \cite{Ye2020}, its computational complexity is in fact $O\left(n^r\left(\log n\right)^r\right)$ and not $O\left(n^r \log n\right)$ as claimed in~\cite{Ye2020}.

\subsection{Existing complexity reduction methods}
\label{sec:reduced_comp}
\subsubsection{Sparse RPA decoder}
The sparse RPA decoder~ (SRPA)\cite{Fathollahi2021} reduced the computational complexity of the RPA decoder by introducing a projection pruning factor $0<q < 1$. Specifically, it keeps $q\times (n-1)$ random projections instead of all $n-1$ projections at each recursive level. The numerical results of this decoder show that removing half of the projections for decoding an $RM(m,2)$ code does not degrade the error-correcting performance significantly. Moreover, for more aggressive pruning, \mbox{k-SRPA} was introduced to improve the decoding performance. It runs $k$ SRPA decoders generating $k$ estimated codewords for a single received vector and then uses a CRC check to choose the best candidate. It utilizes Reed's decoder for k$>2$ to guarantee that the output is always a valid codeword. Numerical results reported in \cite{Fathollahi2021} showed that the computational complexity can be reduced between $50\%$ to $79\%$ compared to RPA. However, having multiple decoders and a Reed decoder for the final decision introduces overhead in terms of hardware implementation. In addition, added CRC bits compromise the effective rate of the RM codes and the randomization of projections introduces additional hardware implementation overhead.

\subsubsection{Reduced complexity RPA decoder}
The work of \cite{JiaJie2021} proposed scheduled RPA, denoted by RPA\textsubscript{SCH}, that reduces the number of projections in successive iterations. RPA\textsubscript{SCH} introduced a decaying parameter $d>1$ defining the number of projections at each iteration $j( j\in [1,N_{\max}])$ as $\lceil \frac{n-1}{d^{j-1}}\rceil$. It keeps all projections for the first iteration since most of the errors get corrected in the first iteration~\cite{HashemipourNazari2021}. After that, it reduces the number of projections by a factor of $d$ at each iteration level. This scheduling technique lowers the computational complexity compared to the baseline RPA by up to $67\%$. However, there is still a lot of room to further reduce the number of projections, especially for RM codes with $r>2$ as we will show in the sequel.

\section{Multi-factor pruned RPA algorithm }
\label{sec:delta_RPA}
Both SRPA and RPA\textsubscript{SCH} significantly reduce the computational complexity as measured by the number of FODs, especially for second-order RM codes. However, the number of FODs is still high for $\text{RM}(m,r>2)$ codes as both methods prune all the recursive levels uniformly. Moreover, RPA\textsubscript{SCH} does not apply the decaying factor for the recursion levels in the iteration layers.
%Moreover, RPA\textsubscript{SCH} \fixme{does not apply the decaying factor for the recursive levels in the iteration layers.}{Can you add another sentence to say what this means? What is the result of this?}
In other words, the decoder being called recursively in RPA\textsubscript{SCH} for decoding $\text{RM}(m,r>2)$ codes starts with all projections even in after the first iteration, resulting in a very large number of FODs for the recursion levels called in each iteration.

In this section, we first describe a general multi-factor projection pruning function and we explain the motivation behind the choice of each factor. We then use this function to introduce a multi-factor pruned RPA (MFP-RPA) algorithm where the pruning factor is a function of both on the iteration and the recursion level, resulting in significantly more aggressive pruning than existing methods while maintaining the error-correcting performance of the original RPA algorithm.
 
\subsection{Projection pruning function}
We define the following pruning function that uses three user-defined factors $\gamma$, $\delta_{\text{itr}}$, and $\delta_{\text{rec}}$, where $0< \gamma, \delta_{\text{itr}}, \delta_{\text{rec}}\leq 1$, and has the iteration number $j$ and the recursion level $l$ as its inputs:
\begin{equation}\label{eq:delta_func}
\Delta(j,l,\gamma,\delta_{\text{itr}},\delta_{\text{rec}})= \gamma \times (\delta_{\text{itr}})^{j-1}\times(\delta_{\text{rec}})^{l-2},
\end{equation}
The output of the pruning function $\Delta(j,l,\gamma,\delta_{\text{itr}},\delta_{\text{rec}})$ is a pruning factor that dictates the fraction of projections that are kept at iteration $j$ for recursion level $l$ compared to RPA decoding. 
%and $\Delta(r,\gamma,\delta_{\text{itr}},\delta_{\text{rec}})\times(n-1) $ is the number of projections for  recursion level $r$ at $j^{\text{th}}(j\in[1,N_{\max}])$ iteration.
These parameters are set off-line.
In the following, we explain the motivation behind each pruning factor $\gamma$, $\delta_{\text{itr}}$, and $\delta_{\text{rec}}$.
\begin{algorithm}[t]
\SetAlgoLined
\textbf{Input: }$\mathbf{L}$, $n$, $r$, $N_{max}$, $\gamma$, $\delta_\text{itr}$, $\delta_{\text{rec}}$\\
\textbf{Output: }Estimated decoded codeword $\mathbf{\hat{c}}$\\

\eIf{$r==1$}{
$\mathbf{\hat{c}}\leftarrow$ FOD($\mathbf{L},n$)

}{
\For{$j=1:N_{max}$}{
 $np=\lceil\Delta(j,r,\gamma,\delta_{\text{itr}},\delta_{\text{rec}})\times(n-1)\rceil$\\
 $\gamma' \leftarrow \gamma\times(\delta_{\text{itr}})^{j-1}$ \label{line:gamma} \\
\For{$t=0:np-1$}{
 $i = t\times\lfloor{\frac{n-1}{np}}\rfloor+1$ \label{line:np}\\
 $\mathbf{L}_{/\mathbb{B}_{i}}\leftarrow$ Projection($\mathbf{L},\mathbb{B}_{i}$)\\
 $\mathbf{\hat{c}}_{/\mathbb{B}_{i}}\leftarrow$ \mbox{MFP-RPA}($\mathbf{L}_{/\mathbb{B}_{i}},\frac{n}{2},r{-}1,N_{max},\gamma',\delta_{\text{itr}},\delta_{\text{rec}}$)\\
$\mathbf{L}_{\text{accu}}\leftarrow$ Aggregation(($\mathbf{L},\mathbf{c}_{/\mathbb{B}_{i}},\mathbb{B}_{i}$)\\
}
$\mathbf{{L}}\leftarrow \frac{\mathbf{L}_{\text{accu}}}{np}$\\
}

 $\mathbf{\hat{c}} \leftarrow \frac{1-\text{sign}(\mathbf{L})}{2}$  \tcp{hard-decision}
}

return $\mathbf{\hat{c}}$\;
\caption{MFP-RPA}
\label{alg:delta}

\end{algorithm}

\subsubsection{Starting pruning factor $\gamma$} 
As mentioned in Section~\ref{sec:RPA}, the RPA decoder starts with $n-1$ projections for decoding an $\text{RM}(m,r)$ code. As \cite{Fathollahi2021} stated, having this number of projections is typically redundant and results in a very large computational complexity. As a result, inspired by \cite{Fathollahi2021}, we define the starting pruning factor $\gamma$, which reduces the number of projections even for the first iteration, in contrast to \cite{JiaJie2021}. Moreover, contrary to the approach taken by \cite{Fathollahi2021}, we allow $\gamma$ to change in each level of recursion as a function of the other pruning factors, as shown in line \ref{line:gamma} of Algorithm~\ref{alg:delta}.

\subsubsection{Iteration pruning factor $\delta_{\text{itr}}$}
Similar to \cite{JiaJie2021}, from our numerical simulation results of RPA, we also concluded that the first iteration is the most crucial one and the importance of further iterations quickly decreases. Therefore, the iteration pruning factor $\delta_{\text{itr}}$ exponentially decreases with the iteration count, as shown in~\eqref{eq:delta_func}.

\subsubsection{Recursion pruning factor $\delta_{\text{rec}}$}
\begin{figure}[!t]
    \centering
    \includegraphics[width=0.27\textwidth]{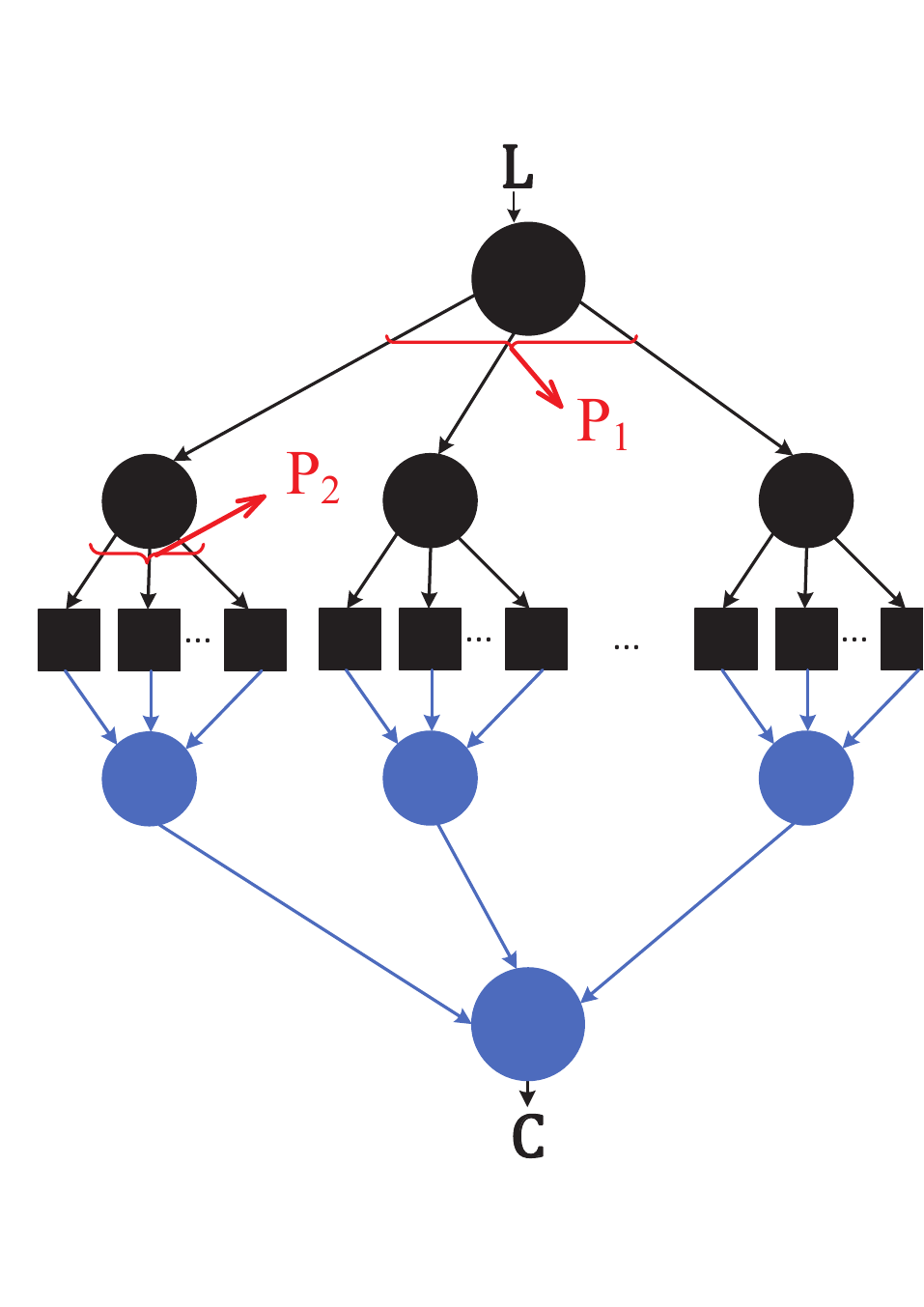}%
    \caption{Pruned version of RPA decoder for an $\text{RM}(m,3)$ codes with $P_1$ and $P_2$ projections at $r=3$ and $r=2$ recursion levels and $N_{\max}=1$.}
    \label{fig:p1p2_bd}
    \vspace*{-1em}
\end{figure}

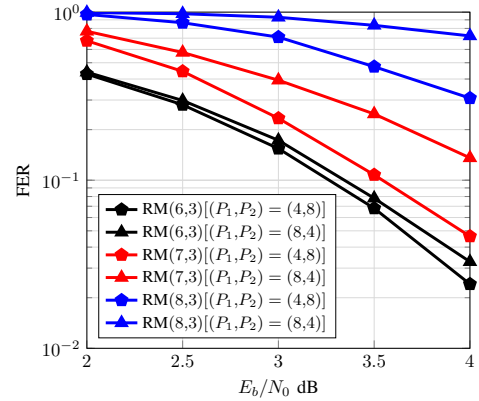
\begin{figure}[!t]
    \centering
    \scalebox{0.7}{%\documentclass[preview]{standalone}
%\usepackage{pgfplots}
%\usepackage[caption=false,font=footnotesize,labelfont=sf,textfont=sf]{subfig}
%\usepgfplotslibrary{groupplots}
%\usepackage{graphicx}
%\usepackage{textcomp}
%\usepackage{xcolor}
%\usepackage{pgfplots, pgfplotstable}
%\usepackage{tikz}
%\usetikzlibrary{matrix, positioning, patterns, shapes, arrows}
%\usetikzlibrary{positioning}
%\usepackage{tikzscale}
%\usepgfplotslibrary{statistics}
%\usepackage{bchart}
%\pgfplotsset{compat=newest}% <- added, current version is 1.13
%\usepgfplotslibrary{colorbrewer}

%% Dummy text
%\usepackage{lipsum}

%% Better looking and squeezes stuff:
%\usepackage{microtype}
%\begin{document}
%%----------------------FER---------------------------------------------
\begin{tikzpicture}
\begin{semilogyaxis}[
    height=0.9 \columnwidth,
    width= \columnwidth,
    yminorgrids = true,
    xminorgrids = true,
	xlabel={$E_b/N_0$~dB},
	ylabel={FER},
	label style={font=\normalsize},
	legend pos=south west,
    ymajorgrids=true,
    xmajorgrids=true,
     every axis plot/.append style={ultra thick},
    xmin = 2,
	xmax = 4,
	xtick={2,2.5,3,3.5,4},
	ymax = 1,
	ymin = 1e-2,
	legend style={font=\small},
	legend cell align={left},
			tick align=inside,
			grid=both, grid style={gray!30},
			/pgfplots/table/ignore chars={|},
%    y tick label style={/pgf/number format/.cd,%          
%          set thousands separator={},
%          fixed},
	 y tick label style={font=\normalsize},
    x tick label style={/pgf/number format/fixed, font=\normalsize},
    			tick align=inside,
			grid=both, grid style={gray!30},
]
%      x tick label style={/pgf/number format/.cd,%
%          scaled x ticks = false,
%          set decimal separator={.},
%          fixed}, 

%--dprunSIPA_floating_tanh_itrs(1)_p1(4)_p2(8)
\addplot[ color=black , mark=pentagon*,mark options={scale=1.5} ] coordinates {
( 2.00, 0.42808219)
( 2.50, 0.28113579)
( 3.00, 0.15429718)
( 3.50, 0.06808742)
( 4.00, 0.02410510)
};

%--hprunSIPA_floating_tanh_itrs(1)_p1(8)_p2(4)
\addplot[ color=black ,mark=triangle*,mark options={scale=1.5} ] coordinates {
( 2.00, 0.43936731)
( 2.50, 0.29806259)
( 3.00, 0.17286085)
( 3.50, 0.07816775)
( 4.00, 0.03274716)
};

%--dprunSIPA_floating_tanh_itrs(1)_p1(4)_p2(8)
\addplot[ color=red ,mark=pentagon*,mark options={scale=1.5} ] coordinates {
( 2.00, 0.67567568)
( 2.50, 0.44444444)
( 3.00, 0.23364486)
( 3.50, 0.10764263)
( 4.00, 0.04640371)
};

%--hprunSIPA_floating_tanh_itrs(1)_p1(8)_p2(4)
\addplot[ color=red ,mark=triangle*,mark options={scale=1.5} ] coordinates {
( 2.00, 0.76923077)
( 2.50, 0.57570524)
( 3.00, 0.39370079)
( 3.50, 0.24789291)
( 4.00, 0.13585111)
};

%--dprunSIPA_floating_tanh_itrs(1)_p1(4)_p2(8)
\addplot[ color=blue ,mark=pentagon*,mark options={scale=1.5} ] coordinates {
( 2.00, 0.96899225)
( 2.50, 0.86505190)
( 3.00, 0.71022727)
( 3.50, 0.47370914)
( 4.00, 0.30797659)
};

%--hprunSIPA_floating_tanh_itrs(1)_p1(8)_p2(4)
\addplot[ color=blue ,mark=triangle*,mark options={scale=1.5} ] coordinates {
( 2.00, 0.99403579)
( 2.50, 0.97847358)
( 3.00, 0.93196645)
( 3.50, 0.83472454)
( 4.00, 0.72202166)
};
\coordinate (top) at (rel axis cs:0,1);
\legend{RM$(6{,} 3)[(P_1{,}P_2)=(4{,}8)]$,RM$(6{,} 3)[(P_1{,}P_2)=(8{,}4)]$,RM$(7{,} 3)[(P_1{,}P_2)=(4{,}8)]$,RM$(7{,} 3)[(P_1{,}P_2)=(8{,}4)]$,RM$(8{,} 3)[(P_1{,}P_2)=(4{,}8)]$,RM$(8{,} 3)[(P_1{,}P_2)=(8{,}4)]$}

\end{semilogyaxis}

\end{tikzpicture}
%%----------------------------------------------------------------------  
%\end{document}
}%
    \caption{Frame error rate (FER) of different $r=3$ RM codes under pruned RPA decoder depicted in Fig.~\ref{fig:p1p2_bd}.}
    \label{fig:p1p2_fer}
    \vspace*{-1em}
\end{figure}

We designed an experiment to explore the impact of each recursive layer on the error-correcting performance of the pruned version of the RPA decoder for various $\text{RM}(m,3)$ codes. We defined two parameters $P_1$ and $P_2$ representing the number of projections in level $r=3$ and $r=2$ for the $\text{RM}(m,3)$ codes, respectively, as shown in Fig.~\ref{fig:p1p2_bd}. The product $P_1\times P_2$ determines the number of required FODs for the generated first-order RM codes. In this experiment, we set $N_{\max}=1$ to remove the effect of iterations on the performance. As shown in Fig.~\ref{fig:p1p2_fer}, with the same overall number of FODs, aggressive pruning in the level $r=2$ leads to a worse error-correcting performance compared to the aggressive pruning in the level $r=3$. As a result, we can conclude that the effect of each recursion level on the overall error-correcting performance of the pruned version of the RPA decoder is clearly not the same. Motivated by this observation, the recursion pruning factor $\delta_{\text{rec}}$ affects the recursion levels such that the highest number of projections is kept at the level $r=2$, meaning that the pruning becomes more aggressive for the higher recursion layer, as~\eqref{eq:delta_func} shows.
%-------------------------------------
%------------------------------------
%-------------------------------------
%------------------------------------
%-------------------------------------
%\begin{figure}[!t]
%\centering
%\subfloat[]{\includegraphics[width=0.22\textwidth]{pruned}%
%\label{fig:p1p2_bd}}
%%\hfil
%\subfloat[]{\includegraphics[width=0.27\textwidth]{p1p2}%
%\label{fig:p1p2_fer}}
%\caption{(a) Pruned version of RPA decoder for an $\text{RM}(m,3)$ codes with $P_1$ and $P_2$ projections at $r=3$ and $r=2$ recursion levels and $N_{\max}=1$. (b) Frame error rate (FER) of different third-order RM codes under pruned RPA decoder depicted in (a).}
%\label{fig:p1p2}
%\end{figure}
%------------------------------------

\subsection{Multi-factor pruned RPA algorithm}
Algorithm \ref{alg:delta} describes the overall proposed multi-factor pruned (MFP-RPA) decoding method for $\text{RM}(m,r)$ codes. Overall, it follows the logic of the baseline RPA decoding algoritum but, as noted in line \ref{line:np}, it selects $np$ projections distributed among all $n-1$ projections uniformly, where $np \ll n-1.$ 
Moreover, SRPA can be obtained using \mbox{MFP-RPA} by setting the tuple $(\gamma,\delta_{\text{itr}},\delta_{\text{rec}})$ to $(q,1,1)$. Similarly, RPA\textsubscript{SCH} can be obtained using \mbox{MFP-RPA} by setting the tuple  $(\gamma,\delta_{\text{itr}},\delta_{\text{rec}})$ to $(1,\frac{1}{d},1)$ for all recursion levels. Finally, the baseline RPA algorithm can also be obtained from \mbox{MFP-RPA} with $(\gamma,\delta_{\text{itr}},\delta_{\text{rec}}) = (1,1,1)$.

\subsection{Complexity of \mbox{MFP-RPA}} 
The number of FODs for \mbox{MFP-RPA} decoding the $\text{RM}(m,r)$ code can be calculated as follows:
\begin{equation}\label{eq:num_FOD}
\lambda_{\text{{MFP-RPA}}} = \sum_{j=1}^{N_{\max}}\prod_{l=2}^{r} \left\lceil\Delta(j,l,\gamma\delta_{\text{itr}}^{j-1},\delta_{\text{itr}},\delta_{\text{rec}}) \times \left(\frac{n}{2^{r-l}}-1\right) \right\rceil.
\end{equation}
%Similar to the previously introduced complexity reduction methods, we consider the number of required FODs as the computational complexity of our proposed \mbox{MFP-RPA}.
Compared to \eqref{eq:rpa_comp}, a good selection of $\gamma$, $\delta_{\text{itr}}$ , and $\delta_{\text{rec}}$ can result in a significant reduction of computational complexity.
This complexity reduction can lead to lower hardware resource requirements, lower latency, or both, depending on the exact hardware architecture that is used (e.g., fully parallel, sequential, or semi-parallel, respectively).
\section{Simulation Results}
\label{sec:results}
%------------------------------------------------------------------
\begin{figure}[t!]
	\centering
	\begin{tikzpicture}
		\begin{groupplot}[group style={group name=fer_queries, group size= 1 by 2, horizontal sep=15pt, vertical sep=42pt},
			footnotesize,
			height=0.62 \columnwidth,
    		width= 0.8\columnwidth,
			xlabel=Eb\slash No dB,%$\frac{E_b}{N_0}$ (dB),
			ymode=log,
			tick align=inside,
			grid=both, grid style={gray!30},
			/pgfplots/table/ignore chars={|},
			every axis plot/.append style={ ultra thick},
			ylabel={FER},
			label style={font=\normalsize},
			ymin=1e-5, ymax = 1e-1, 
			]

%-- RM(2,7) ------------------------------------------
			\nextgroupplot[ytick pos=left,xmin=1.5,xmax=4, xtick={1.5,2,2.5,3,3.5,4}]
			
%--RPA_floating_tanh_itrs(3)
\addplot[ color=red ,mark=* ] coordinates {
( 1.50, 0.02659574)
( 2.00, 0.00887367)
( 2.50, 0.00230363)
( 3.00, 0.00053300)
( 3.50, 0.00009200)
( 4.00, 0.00001100)
};

%--deltaRPA_floating_tanh_itrs(3)_g(0.67)_dI(0.25)_dR(0.50)
\addplot[ color=blue ,mark=diamond*,mark options={scale=1.5} ] coordinates {
( 1.50, 0.02949504)
( 2.00, 0.00968607)
( 2.50, 0.00267543)
( 3.00, 0.00059600)
( 3.50, 0.00010200)
( 4.00, 0.00001300)
};\label{gp:deltaRPARM27}

%%--2Spars_RPA_floating_tanh_itrs(3)_d=2
\addplot[ color=black ,mark=triangle* ] coordinates {
%( 1.00, 0.08)
%( 1.50, 0.033)
( 2.00, 0.012)
( 2.50, 0.0035)
( 3.00, 0.00075)
( 3.50, 0.0001500)
%( 4.00, 0.00001500)
};\label{gp:SRPA}

%%--reduced_RPA_floating_tanh_itrs(3)_d=2
\addplot[ color=brown ,mark=asterisk ] coordinates {
( 1.50, 0.028)
( 2.00, 0.009)
( 2.50, 0.0025)
( 3.00, 0.0006)
( 3.50, 0.0001)
( 4.00, 0.000012)
};\label{gp:RPASCH}

%\legend{RPA{:}$RM(2{,} 7)$,IPA{:}$RM(2{,} 7)$}

			\coordinate (top) at (rel axis cs:0,1);
%-- RM(3,8) ------------------------------------------
		
\nextgroupplot[xmin=1, xmax=3, xtick={1,1.5,2,2.5,3},legend style={nodes={scale=0.75, transform shape}},legend cell align={left},legend pos=south west]
			\coordinate (bot) at (rel axis cs:1,0);

%--RPAp_floating_tanh_itrs(3)
\addplot[ color=red ,mark=* ] coordinates {
( 1.00, 0.07665185)
( 1.50, 0.01678585)
( 2.00, 0.00260281)
( 2.50, 0.00024600)
( 3.00, 0.00001200)
};\label{gp:RPA}

%--deltaRPA_floating_tanh_itrs(3)_g(0.75)_dI(0.33)_dR(0.75)
\addplot[ color=blue ,mark=pentagon* ] coordinates {
( 1.00, 0.08836264)
( 1.50, 0.01995729)
( 2.00, 0.00316350)
( 2.50, 0.00028400)
( 3.00, 0.00001600)
};\label{gp:deltaRPARM38}

%%--reduced_RPA_floating_tanh_itrs(3)_d=2
\addplot[ color=brown ,mark=asterisk ] coordinates {
( 1.00, 0.09)
( 1.50, 0.02)
( 2.00, 0.0029)
( 2.50, 0.00031)
( 3.00, 0.0000250)
};

%%--SRPA_floating_tanh_itrs(3)_1/8
\addplot[ color=black ,mark=triangle* ] coordinates {
( 1.00, 0.095)
( 1.50, 0.027)
( 2.00, 0.005)
( 2.50, 0.0005)
%( 3.00, 0.00004)
};

%\legend{RPA{:}$RM(3{,} 8)$,RPA\textsubscript{delta}{:}$RM(3{,} 8)$}

%\legend{RPA floating (tanh),RPA floating (MinSum),SIPA floating,SIPA Quant($3{:}2$),SIPA Quant($3{:}3$),SIPA Quant($3{:}4$)}
		\end{groupplot}
		\node[below = 0.8cm of fer_queries c1r1.south] { (a){\footnotesize $RM(7,2)$} };
		\node[below = 0.8cm of fer_queries c1r2.south] {(b){\footnotesize $RM(8,3)$}};
%		\node[below = 1.2cm of fer_queries c1r1.south] {\footnotesize $RM(7,2)$};
%		\node[below = 1.2cm of fer_queries c2r1.south] {\footnotesize $RM(8,3)$};		

		\path (top|-current bounding box.north) -- coordinate(legendpos) (bot|-current bounding box.north);
		\matrix[
		matrix of nodes,
		anchor=south,
		draw,
		inner sep=0.1em,
		draw,
		column 1/.style={anchor=base west},
    	column 2/.style={anchor=base west},
    	column 3/.style={anchor=base west},
    	column 4/.style={anchor=base west},
%    	column 5/.style={anchor=base west},
%    	column 6/.style={anchor=base west}, 
%    	column 7/.style={anchor=base west},
%    	column 8/.style={anchor=base west},
		]at(legendpos)
		{
			\ref{gp:RPA}& \small RPA  &[3pt]
			\ref{gp:SRPA}& \small $2$-SRPA($q=\frac{1}{8}$) \\
			\ref{gp:RPASCH}& \small RPA\textsubscript{SCH}($d=2$) &[3pt]
			\ref{gp:deltaRPARM27}& \small MFP-RPA{$(\frac{2}{3},\frac{1}{4},\frac{1}{2})$}  \\
			\ref{gp:deltaRPARM38}& \small MFP-RPA{$(\frac{3}{4},\frac{1}{3},\frac{3}{4})$} \\
			}; 
	\end{tikzpicture}
%%----------------------------------------------------------------------  
	\caption{FER comparison between RPA, $2$-SRPA, RPA\textsubscript{SCH}, and \mbox{MFP-RPA} algorithms. The results for $2$-SRPA and RPA\textsubscript{SCH} are taken from \cite{Fathollahi2021} and \cite{JiaJie2021}, respectively.}
	\label{fig:FER}
\end{figure}
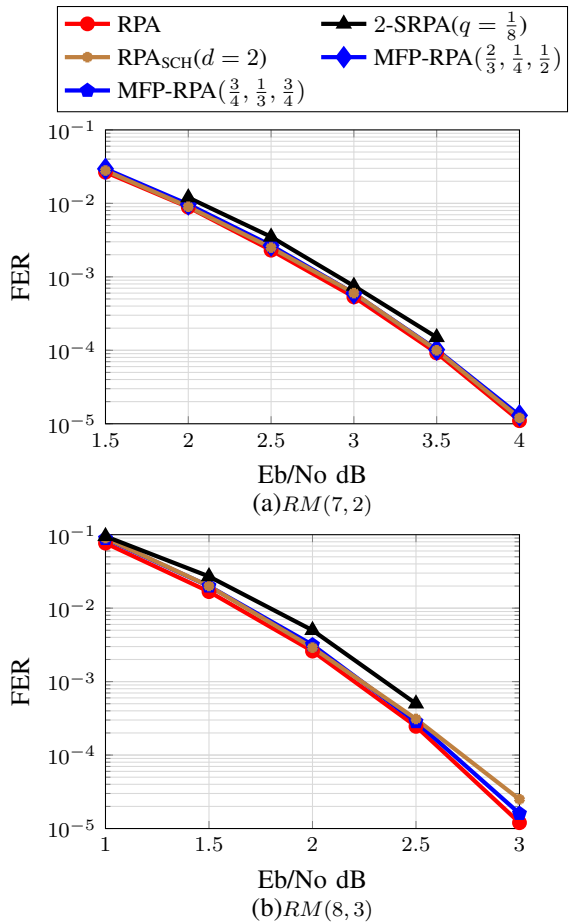
%-----------------------------------------------------------------------
We simulate the performance of the proposed \mbox{MFP-RPA} algorithm on $\text{RM}(7,2)$ and $\text{RM}(8,3)$ codes over the additive white Gaussian noise (AWGN) channel. 
Fig.~\ref{fig:FER} shows the simulation results for the proposed \mbox{MFP-RPA} compared to RPA~\cite{Ye2020}, SRPA~\cite{Fathollahi2021}, and RPA\textsubscript{SCH}~\cite{JiaJie2021}. 
%We set $N_{\max}$ to $\lceil\frac{m}{2}\rceil$ as in~\cite{Abbe2020}, which is $N_{\max}=4$ for both $\text{RM}(7,2)$ and $\text{RM}(8,3)$ codes.
From our simulations with $N_{\max}=\lceil\frac{m}{2}\rceil$, we observed that the last iteration does not impact the error-correcting performance. Therefore, we set the maximum number of iterations $N_{\max}=3$ for RPA and \mbox{MFP-RPA}. 
Moreover, we set the tuple $(\gamma,\delta_{\text{itr}},\delta_{\text{rec}})$ of user-defined pruning factors of \eqref{eq:delta_func} to  $\left(\frac{2}{3},\frac{1}{4},\frac{1}{2}\right)$ for the $\text{RM}(7,2)$ code and $\left(\frac{3}{4},\frac{1}{3},\frac{3}{4}\right)$ for the $\text{RM}(8,3)$ code. 
We note that we selected the pruning factors heuristically in a way that the performance loss compared to the non-pruned decoder is negligible and we currently do not have a systematic way of choosing these factors. Furthermore, the code's parameters as well as required reliability, latency, and available hardware resources in the event of hardware implementation should generally be taken into consideration for selecting the appropriate pruning factors.

%It should be pointed out that for $2$-SRPA decoding of $\text{RM}(8,3)$, two SRPA decoders are used in the first level of recursion, i.e., $r=3$. However, four SRPA decoders are used when the recursions reach $\text{RM}(7,2)$.
%In addition, the results shown in Fig.~\ref{fig:FER} for $2$-SRPA are taken from \cite{Fathollahi2021} with $N_{\max}=\lfloor\frac{m}{2} \rfloor$.
We observe that the proposed \mbox{MFP-RPA} algorithm has effectively identical error-correcting performance to the RPA algorithm and its previously proposed pruning methods SRPA and RPA\textsubscript{SCH}. However, as shown in Table ~\ref{table:fht}, it reduces the number of FODs significantly. Specifically, Table ~\ref{table:fht} illustrates the number of FODs required for the baseline RPA, RPA\textsubscript{SCH}, $2$-SRPA, and our proposed \mbox{MFP-RPA}. 
We note that the numbers in Table~\ref{table:fht} do not use early stopping controlled by $\theta$ in~\eqref{eq:earlystop} and we use the worst-case complexity numbers for SRPA and RPA\textsubscript{SCH} as a hardware implementation generally has to account for the worst case. However, early stopping can still be used in conjunction with \mbox{MFP-RPA} if desired to, e.g., reduce the energy consumption.
%%----------------------------------------------------------------------
\begin{table}[t!]
\caption{Comparison the number of FODs required for RPA~\cite{Ye2020}, RPA\textsubscript{SCH}~\cite{JiaJie2021}, 2-SRPA~\cite{Fathollahi2021}, and our proposed \mbox{MFP-RPA}.}

\centering
\resizebox{\columnwidth}{!}{%
\begin{tabular}{lccccc}
\hline
\multirow{3}{*}{\textbf{}} & \multirow{3}{*}{RPA} & \multirow{3}{*}{RPA\textsubscript{SCH}} & \multirow{3}{*}{$2$-SRPA} & \multicolumn{2}{c}{\mbox{MFP-RPA}$(\gamma,\delta_{\text{itr}},\delta_{\text{rec}})$}                    \\
                           &                      &                       &                       & \multicolumn{1}{c}{$(2/3,1/4,1/2)$} & \multicolumn{1}{c}{$(3/4,1/3,3/4)$} \\ \hline
 $\text{RM}(7,2)$ &    $381$            &      $221$             &    $96$             &   $113$              &                     -  \\ \hline
 $\text{RM}(8,3)$  &  $291465$            &    $98385$          &    $36433$       &        -              &                  $22544$ \\ \hline
\end{tabular}
}
\label{table:fht}
\end{table}
%----------------------------
%\fixme{Moreover, for the fair comparison, we considered $N_{\max}=3$ for all algorithms mentioned in Table~\ref{table:fht}.}{This is the third time you mention $N_{\max}$.You should say it once, pick a value and stick with it, otherwise you confuse the reader.} 
We observe that, for the $\text{RM}(7,2)$ code, our proposed MFP-RPA reduces the number of FODs by $70$\% and $49$\% compared to the RPA and RPA\textsubscript{SCH} algorithms. Compared to 2-SRPA, MFP-RPA requires $18$\% more FODs, but it also has a better error-correcting performance. For the $\text{RM}(8,3)$ code, MFP-RPA reduces the number of FODs by $92$\%, $77$\%, and $38$\% compared to RPA, RPA\textsubscript{SCH}, and 2-SRPA, respectively.
We proposed the update rule \eqref{eq:delta_func}  based on the intuitions explained in Section~\ref{sec:delta_RPA}. However, other update rules are possible and they may lead to further complexity reduction. This is an interesting open problem.
%Therefore, for RM codes with order$>2$, our proposed recursion pruning factor impacts the computational complexity remarkably.

%reducing the required number of FHTs by $70\%$ without loosing the error-correcting performance.

\section{Conclusion}
\label{sec:conclusion}

%The high computational complexity of RPA decoding due to the very large number of projections, makes it infeasible to decode RM codes with order $r>2$ in practice, especially for time-critical applications with limited hardware resource budgets. The previously proposed methods to lower the computational complexity of the RPA decoder also introduced the overhead in required resources, making the implementation of the RPA decoder challenging. 
In this work, we proposed a multi-factor pruning method for RPA decoding of RM codes that prunes projections as a function of both the iteration number and the recursion level. Our results show that significantly more aggressive projection pruning is possible compared to existing methods without degrading the error-correcting performance. Specifically, our proposed multi-factor pruning method leads to up to $92\%$ and $77\%$ lower computational complexity compared to the baseline RPA decoding algorithm and previously proposed complexity reduction techniques, respectively.

%\newpage
\balance
\bibliographystyle{ieeetr}
\bibliography{refs}

\begin{thebibliography}{10}

\bibitem{Muller1954}
D.~E. Muller, ``Application of boolean algebra to switching circuit design and
  to error detection,'' {\em Transactions of the IRE Professional Group on
  Electronic Computers}, vol.~{EC}-3, pp.~6--12, Sep 1954.

\bibitem{Reed1954}
I.~Reed, ``A class of multiple-error-correcting codes and the decoding
  scheme,'' {\em Transactions of the {IRE} Professional Group on Information
  Theory}, vol.~4, pp.~38--49, Sep 1954.

\bibitem{Betextquotesingleery1986}
Y.~Be{\textquotesingle}ery and J.~Snyders, ``Optimal soft decision block
  decoders based on fast {Hadamard} transform,'' {\em {IEEE} Transactions on
  Information Theory}, vol.~32, pp.~355--364, May 1986.

\bibitem{Ashikhmin1996}
A.~E. Ashikhmin and S.~N. Litsyn, ``Fast decoding algorithms for first order
  {Reed{\textendash}Muller} and related codes,'' {\em Designs, Codes and
  Cryptography}, vol.~7, pp.~187--214, Mar 1996.

\bibitem{green1966serial}
R.~Green, ``A serial orthogonal decoder,'' {\em Jet Propulsion Laboratory (JPL)
  Space Programs Summary}, vol.~37, pp.~247--253, 1966.

\bibitem{Sidel1992}
V.~M. Sidel'nikov and A.~S. Pershakov, ``Decoding of {Reed{\textendash}Muller}
  codes with a large number of errors,'' {\em Problemy Peredachi Informatsii},
  vol.~28, pp.~80--94, 1992.

\bibitem{Dumer2004}
I.~Dumer, ``Recursive decoding and its performance for low-rate
  {Reed{\textendash}Muller} codes,'' {\em {IEEE} Transactions on Information
  Theory}, vol.~50, pp.~811--823, May 2004.

\bibitem{Sakkour2005}
B.~Sakkour, ``Decoding of second order {Reed{\textendash}Muller} codes with a
  large number of errors,'' in {\em {IEEE} Information Theory Workshop}, Aug.
  2005.

\bibitem{Dumer2006a}
I.~Dumer, ``Soft-decision decoding of {Reed{\textendash}Muller} codes: a
  simplified algorithm,'' {\em {IEEE} Transactions on Information Theory},
  vol.~52, pp.~954--963, Mar 2006.

\bibitem{Dumer2006}
I.~Dumer and K.~Shabunov, ``Soft-decision decoding of {Reed{\textendash}Muller}
  codes: {Recursive lists},'' {\em {IEEE} Transactions on Information Theory},
  vol.~52, pp.~1260--1266, Mar. 2006.

\bibitem{Arikan2009}
E.~Arikan, ``Channel polarization: A method for constructing capacity-achieving
  codes for symmetric binary-input memoryless channels,'' {\em {IEEE}
  Transactions on Information Theory}, vol.~55, pp.~3051--3073, Jul 2009.

\bibitem{Hussami2009}
N.~Hussami, S.~B. Korada, and R.~Urbanke, ``Performance of polar codes for
  channel and source coding,'' in {\em {IEEE} International Symposium on
  Information Theory}, Jun 2009.

\bibitem{Kudekar2017}
S.~Kudekar, S.~Kumar, M.~Mondelli, H.~D. Pfister, E.~Sasoglu, and R.~L.
  Urbanke, ``{Reed{\textendash}Muller} codes achieve capacity on erasure
  channels,'' {\em {IEEE} Transactions on Information Theory}, vol.~63,
  pp.~4298--4316, Jul 2017.

\bibitem{Abbe2015}
E.~Abbe, A.~Shpilka, and A.~Wigderson, ``{Reed{\textendash}Muller} codes for
  random erasures and errors,'' in {\em Proceedings of the forty-seventh annual
  {ACM} symposium on Theory of Computing}, {ACM}, Jun 2015.

\bibitem{Tonnellier2021}
T.~Tonnellier, M.~Hashemipour-Nazari, N.~Doan, W.~J. Gross, and
  A.~Balatsoukas-Stimming, ``Towards practical near-maximum-likelihood decoding
  of error-correcting codes: An overview,'' in {\em {IEEE} International
  Conference on Acoustics, Speech and Signal Processing}, Jun 2021.

\bibitem{Tal2011}
I.~Tal and A.~Vardy, ``List decoding of polar codes,'' in {\em {IEEE}
  International Symposium on Information Theory Proceedings}, Jul 2011.

\bibitem{Ye2020}
M.~Ye and E.~Abbe, ``Recursive projection-aggregation decoding of
  {Reed{\textendash}Muller} codes,'' {\em {IEEE} Transactions on Information
  Theory}, vol.~66, pp.~4948--4965, Aug. 2020.

\bibitem{Lian2020}
M.~Lian, C.~Hager, and H.~D. Pfister, ``Decoding {Reed{\textendash}Muller}
  codes using redundant code constraints,'' in {\em {IEEE} International
  Symposium on Information Theory}, Jun 2020.

\bibitem{Fathollahi2021}
D.~Fathollahi, N.~Farsad, S.~A. Hashemi, and M.~Mondelli, ``Sparse
  multi-decoder recursive projection aggregation for {Reed{\textendash}Muller}
  codes,'' in {\em {IEEE} International Symposium on Information Theory}, Jul
  2021.

\bibitem{JiaJie2021}
J.~Li, S.~M. Abbas, T.~Tonnellier, and W.~J. Gross, ``Reduced complexity {RPA}
  decoder for {Reed{\textendash}Muller} codes,'' in {\em {IEEE} International
  Symposium on Topics in Coding}, Sep 2021.

\bibitem{Abbe2020}
E.~Abbe and M.~Ye, ``{Reed{\textendash}Muller} codes polarize,'' {\em {IEEE}
  Transactions on Information Theory}, vol.~66, pp.~7311--7332, Dec 2020.

\bibitem{HashemipourNazari2021}
M.~Hashemipour-Nazari, K.~Goossens, and A.~Balatsoukas-Stimming, ``Hardware
  implementation of iterative projection-aggregation decoding of
  {Reed{\textendash}Muller} codes,'' in {\em {IEEE} International Conference on
  Acoustics, Speech and Signal Processing}, Jun 2021.

\end{thebibliography}
\vfill

\end{document}